# First-Principles Theory of Competing Order Types, Phase Separation, and Phonon Scattering in Thermoelectric Pb-Sb-Ag-Te Alloys

S. V. Barabash,[†] V. Ozolins,[†] and C. Wolverton[§]

[†]*Department of Materials Science and Engineering, University of California, Los Angeles, California 90095-1595*

[§] *Department of Materials Science & Engineering, Northwestern University, Evanston, Illinois 60208*

(Date: February 27, 2008)

Using a first-principles cluster expansion, we shed light on the solid-state phase diagram and structure of the recently discovered high-performance thermoelectrics, Pb-Ag-Sb-Te alloys. The observed nanoscale inhomogeneities are shown to be coherent precipitates of ordered $Ag_mSb_nTe_{m+n}$ phases, such as $AgSbTe_2$, all immiscible with rocksalt PbTe. The solubility is high for PbTe in $AgSbTe_2$ and low for (Ag,Sb)Te in PbTe (8% vs. 0.6% at 850 K). The differences in the phonon spectra of PbTe and $AgSbTe_2$ confirm that the inhomogeneities enhance the thermoelectric performance.

Recent measurements[1] of an exceptionally high thermoelectric figure of merit $Z$ ($ZT$~2 at T=800 K) in lead-antimony-silver-telluride (LAST) alloys $Pb_{2-x-y}Ag_ySb_xTe_2$ have turned these materials into promising candidates for use in thermoelectric power generation devices. The mechanism of the increase in $ZT$ beyond the values achieved in the pure constituents (ZT~0.8 in PbTe [2] and ZT<1.4 $AgSbTe_2$ [3]) is not yet clear; it has been suggested[1] to originate from strong compositional fluctuations across the specimens, which could lead to an additional phonon scattering, and thereby to the observed suppression[3,4] of thermal conductivity. The presence of inhomogeneities in high-$Z$ LAST alloys has been demonstrated in several recent studies.[1,5,6,7] However, currently there is no consensus on the atomic structure of LAST materials, much less on the physical driving force for the formation of the observed inhomogeneities and possible mechanisms of enhanced phonon scattering.

In fact, even the structure of a pure $AgSbTe_2$ compound is not reliably established: long thought to have a cubic rocksalt structure with *random* occupation of the cation sublattice[8], $AgSbTe_2$ was recently demonstrated to exhibit *ordering* of Ag and Sb ions[6], but the type of ordering remains elusive[9]. The structure of alloys between $AgSbTe_2$ and PbTe is even less understood. Most early studies reported that miscible disordered solid solutions exist in the entire concentration range[10,11,12], although a miscibility gap was also reported[13]. Recent experiments have uncovered strong evidence that many alloys that appear homogeneous in powder X-ray and optical microscopy studies are in fact composed of nano- to micron-scale regions with *varying* composition[7], atomic ordering[6], thermoelectric properties[5], and lattice parameter[6].

On the theoretical side, the largest breakthrough has been the prediction[14] of the likely ordered structures of $AgSbTe_2$. However, only four pre-selected ordered $AgSbTe_2$ structures were considered[14], leaving open the possibility that the true ground state might have been missed. Moreover, it was not established whether the atomic order survives at the typical processing temperatures (>1000 K). In other studies, ordering tendencies in Pb-containing LAST alloys have only been analyzed using an *ad-hoc* model Hamiltonian[15], or assuming that Ag and Sb are present as dilute impurities in PbTe[16,17]. First-principle studies at higher Ag and Sb concentrations have not been attempted, and the composition-temperature phase diagram and phase stability of LAST alloys remain poorly understood.

Here, we report an exhaustive first-principles study of the microstructure and thermodynamics of ordering and phase-separation in the quasi-ternary $Pb_{1-x-y}Ag_xSb_yTe$ system. We use the cluster expansion method to identify the T=0 K stable phases by explicitly searching through $\sim 10^5$ quaternary structures. The calculated composition-temperature phase diagram for the coherent[18] $(PbTe)_{1-2x}(AgSbTe_2)_x$ system confirms the existence of a miscibility gap, suggesting that the observed compositional inhomogeneities in LAST thermoelectrics are due to precipitation of ordered $AgSbTe_2$-rich precipitates from the PbTe-rich solid solution. The calculated ordering temperature of the $AgSbTe_2$ phase is 1150±50 K, showing that $AgSbTe_2$ precipitates stay ordered up to the melting point of the material. While PbTe tolerates only a small degree of substitution by Ag and Sb on the Pb sites (up to 0.6% at 850 K), the solid solubility of Pb in the ordered $AgSbTe_2$ phase can be as high as 8% at 850 K, with Pb predominantly occupying sites on the Sb sublattice. The ordering temperature of $AgSbTe_2$ is insensitive to the Pb content. Finally, we demonstrate that there are substantial differences in the low-energy region of the phonon spectra of PbTe and $AgSbTe_2$, which are expected to enhance acoustic phonon scattering at matrix/precipitate interfaces and suppress thermal conductivity.

The multicomponent cluster expansion (CE) approach[19] was used to construct a generalized Ising-like Hamiltonian for ternary rocksalt-based $Pb_{1-x-y}Ag_xSb_yTe$ structures, where the cation sublattice of the rocksalt lattice can be occupied by Pb, Ag and Sb atoms. We used the ATAT toolkit[20] to obtain the optimal set of pair and many-body effective cluster interactions, starting from fully relaxed total energies of 86 ordered input structures. The total energies of these structures were calculated within the density-functional theory[21,22] using projector-augmented wave potentials[23] as implemented in the VASP code[24]. Vibrational spectra were calculated using the supercell frozen phonon approach.[25] The resulting CE Hamiltonian had the prediction accuracy of 6% on the scale of the energy difference between the most and the least stable structures[26]. The finite-temperature results were obtained using Metropolis Monte Carlo routines from ATAT.

We considered the energetics of both coherent phases (i.e., those based on the rocksalt lattice) and the known incoherent Ag-Te and Sb-Te compounds. In Fig. 1a, we show the CE-predicted T=0 K formation enthalpies of all ~$10^5$ rocksalt-based ordered structures with less than 20 atoms per unit cell. By constructing a "convex hull" (edges shown as black lines) such that no structure lies below the planes connecting the vertices of this object, we can identify the stable T=0 K "coherent ground state" structures. The energies of all other rocksalt-based structures lie above the "convex hull"; thus, all those other structures may lower their energy by phase separating into some of the "coherent ground state" structures. Fig. 1a shows that the coherent phase stability within the rocksalt-based AgTe-SbTe alloy system is compound-forming and includes several ordered structures[27] besides $AgSbTe_2$, and that all the *quaternary* compounds in the PbTe-(Ag,Sb)Te system are *unstable* with respect to phase separation into PbTe and one of the (Ag,Sb)Te compounds. From the latter observation, we conclude that the observed inhomogeneities[1,5,6,7] in (Pb,Ag,Sb)Te alloys are coherent nanoscale precipitates of immiscible (Ag,Sb)Te-rich phases.

In Fig. 1b we compare the formation energies of the rocksalt-based structures with those of the other known (Ag,Sb)Te phases. We see that at low temperatures all the rocksalt-based (Ag,Sb)Te structures discussed in the previous paragraph are, in fact, unstable with respect to phase separation into incoherent non-rocksalt based phases. However, the energy differences between coherent rocksalt-based and incoherent non-rocksalt-based phases are very small (~5 meV/atom) for all compounds except SbTe. Such small energy differences might explain why contradictory experimental assessments have been made[12,28,29] for the relative stability of $AgSbTe_2$ and the equimolar mixture of $Ag_2Te$ and $Sb_2Te_3$. Most importantly, (Ag,Sb)Te-rich coherent *precipitates* inside the PbTe matrix will be constrained to the rocksalt lattice, leaving $AgSbTe_2$ as the only possible candidate, irrespective of the relative stability of *bulk* $AgSbTe_2$ vs. $Ag_2Te + Sb_2Te_3$. Among the ground state structures of Fig. 1, energetically the most prominent is $AgSbTe_2$. We find that the lowest energy $AgSbTe_2$ is cubic D4, however, it is separated from another structure, trigonal L1$_1$, by only 1.7meV/cation. L1$_1$ and D4 (visualized in Fig. 1c) were among the four structures considered earlier by Hoang *et al.*[14]; here we explicitly confirm that these nearly degenerate structures have the lowest energy among ~10$^5$ possible rocksalt-based AgSbTe$_2$ structures. The coherency strain with a surrounding PbTe matrix does not affect L1$_1$ vs. D4 energetics: restricting the c/a ratio of the L1$_1$ structure to the ideal, cubic rocksalt value and distorting both L1$_1$ and D4 to the lattice constant of PbTe still yields the same 1.7 meV difference in favor of D4.

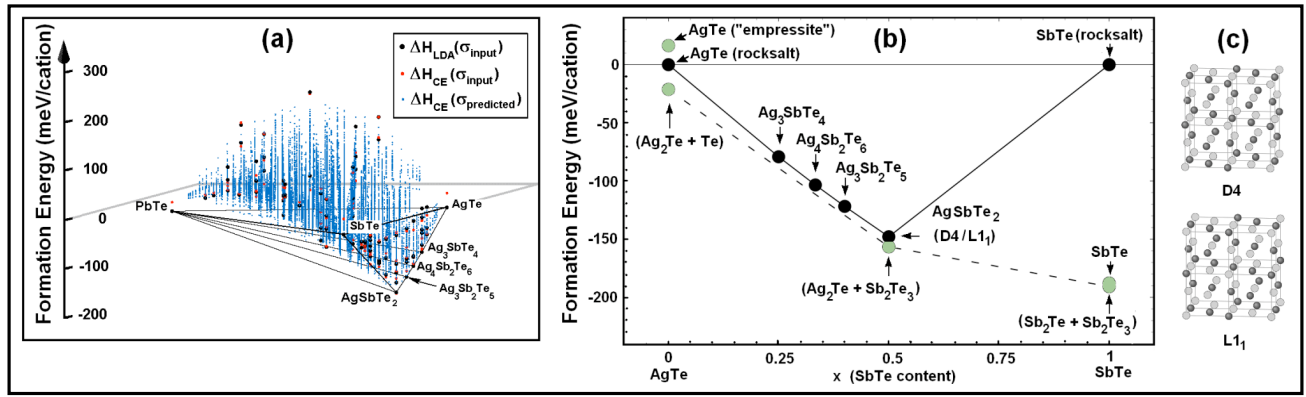

**Figure 1. (color online).** (a): The LDA and CE T=0 K formation enthalpies of the 86 "input" (Ag,Pb,Sb)Te structures (large black and red points, respectively). Small blue points represent the CE enthalpies of ~$10^5$ candidate quaternary structures with ≤20 atoms. Enthalpies are given relative to rocksalt PbTe, AgTe, and SbTe. The black lines highlight the convex hull connecting the "ground state structures" stable within the rocksalt lattice. (b) LDA formation enthalpies of AgTe-SbTe phases at T=0 K. Black dots denote the rocksalt-based structures, green dots denote competing non-rocksalt phases and two-phase mixtures. (c) D4 and $L1_1$ ordering of Ag and Sb sublattice in $AgSbTe_2$ (for visual clarity, Te sublattice is not shown).

The near-degeneracy of L1$_1$ and D4 has a simple geometric reason: it can be shown[30] that these two structures have identical pair- and three-body correlations for *any* neighbor shell. The smallest cluster that distinguishes L1$_1$ from D4 is the four-body nearest-cation-neighbor tetrahedron: L1$_1$ is comprised of Ag$_3$Sb and AgSb$_3$, and D4 – of Ag$_4$, Sb$_4$ and Ag$_2$Sb$_2$ tetrahedra. Experimentally, L1$_1$ and D4 can be distinguished on the basis of a trigonal distortion[31] or by a direction-sensitive probe such as single-crystal X-ray diffraction[32].

In Fig. 2, we show the coherent[18] phase diagram of PbTe-AgSbTe$_2$ solid solutions, as obtained from Monte Carlo simulations[33]. It shows that the ordering tendencies in the AgSbTe$_2$ phase are very strong, since the calculated order-disorder transition temperature $T_{ord}$ in the solid state exceeds the experimentally measured melting temperature of pure AgSbTe$_2$. Furthermore, it is evident that the PbTe phase can accommodate very little substitution by Ag and Sb (only about 0.6% at the binodal equilibrium at 850 K, or about 1% at 1000 K); on the contrary, the AgSbTe$_2$ phase accommodates plenty of Pb substitution (7.5±3% at 850 K). Remarkably, this high degree of Pb substitution does not noticeably affect the calculated ordering temperature, which is predicted to remain practically unchanged upon increasing Pb content.

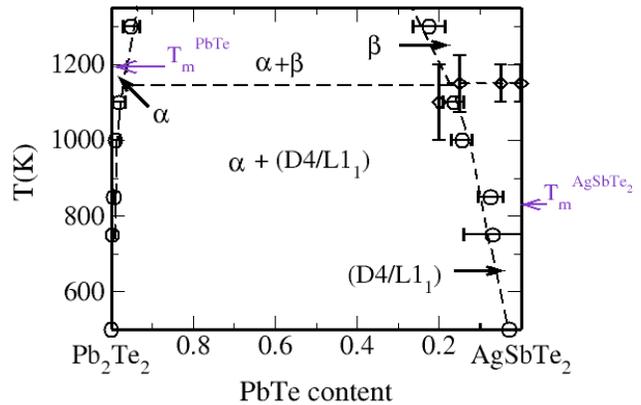

**Figure 2. (color online)** Calculated coherent rocksalt-based solid-state phase diagram of (PbTe)$_{1-x}$(AgSbTe$_2$)$_{x/2}$, showing the boundaries of the immiscible disordered α and β solid solution phases (circles) and the order-disorder transition temperature $T_{ord}$ for the (D4/L1$_1$) ordered phase (diamonds). Also marked are the experimental melting temperatures $T_m$ of pure bulk AgSbTe$_2$ and PbTe. The error bars show the MC uncertainty. All dashed lines are guides for eye.

We also marked in Fig. 2 the experimental melting temperatures $T_m$ of *pure bulk* PbTe and AgSbTe$_2$. An interesting possibility is that in the temperature region between the both bulk $T_m$, AgSbTe$_2$ precipitates may melt inside a solid PbTe matrix, especially for larger sizes. Similar phenomena have been observed for small precipitates in metallic alloys, such as Pb in Al[34]. The existence of the disordered solid phase β is highly hypothetical, although not entirely implausible: it requires that the melting temperatrue of the AgSbTe$_2$ precipitates is sufficiently raised above the value of the pure bulk material $T_m$(AgSbTe$_2$) by Pb doping and/or finitie-size effects. We hypothesize that heat treatment between $T_m$(AgSbTe$_2$) and $T_m$(PbTe) could provide additional means of manipulating the microstructure and ordering of inhomogeneous regions in LAST alloys.

Figure 3a shows a Monte Carlo snapshot of the shape of a AgSbTe$_2$ precipitate in PbTe matrix at 800 K. It is seen that even at these small sizes, distinct facets can be observed, with the most prominent being (111) facets.

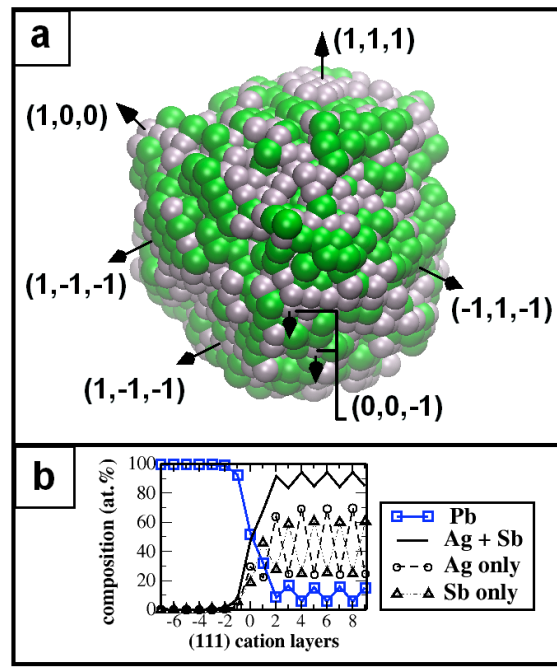

**Figure 3 (color online).** (a) Precipitate of a D4 AgSbTe$_2$ inside PbTe matrix obtained in Monte-Carlo simulation at 800K. For clarity, only Ag (silver) and Sb (green) atoms of the precipitate phase are shown. (b) Variation of the atomic composition of (111) cation layers in the vicinity of a (111) boundary between PbTe and D4 AgSbTe$_2$.

Figure 3b shows the calculated composition profile across a PbTe/(AgSbPb)Te (111) interface. One can see that the width of this interface is 3-4 cationic layers, suggesting that this interface should be classified as diffuse and entropic contributions to the interfacial free energy are not small. We note here that in our calculations the precipitate shape is determined only by the interfacial energies and the effects of elastic stresses are not included. The latter are expected to become increasingly important with increasing precipitate size. A parameter characterizing the ratio of the interfacial and elastic energies, $L=l\varepsilon^2 C_{44}/\sigma$ can be used to estimate sizes where elastic effects start to dominate[35]; here $l$ is the characteristic size of the precipitate, $\varepsilon$ is a size misfit between PbTe and (AgSbPb)Te phases, $C_{44}$ is one of the elastic constants, and $\sigma \sim 3\times 10^3 (mJ/m^2)$ is our estimated interfacial energy between PbTe and $AgSbTe_2$. Using the experimental PbTe elastic constant[36] and the calculated lattice mismatch and the interfacial energies, we obtain that this characteristic size is approximately 30 nm.

Analyzing the atomic structures obtained during MC simulated annealing, we find that D4 forms in the majority of the runs and is thus strongly preferred over $L1_1$[37], despite the small energy difference between the two structures (less than 2% of $T/k_B$ per cation at the ordering transition). Next, we analyzed the atomic order in the partially ordered structures at T=750 K. We found that at the pure $Ag_{0.5}Sb_{0.5}Te$ composition, both Ag and Sb atoms occupy almost exclusively their respective sublattices. Upon a substitution of 11% of cations by Pb atoms, Ag atoms were found to remain constrained to the Ag-rich sublattice, whereas a substantial fraction of Sb atoms (11%) move to the Ag-rich sublattice, forming Sb antisite defects (see Fig. 3b)[33]. At the same time, Pb atoms themselves have a strong preference for selectively occupying the Sb-rich sublattice (in fact, only 8% of all Pb atoms occupy the Ag-rich sublattice). Thus, the $Ag_{0.45}Sb_{0.44}Pb_{0.11}Te$ alloy behaves as a hypothetical $Ag_{0.45}X_{0.55}Te$ with *nearly perfect* order possible at that composition. In this sense, Pb atoms do *not* act to suppress the existing ordering, which explains the result that the ordering temperature remains practically unchanged by Pb substitution (see Fig. 2).

The calculated phonon densities-of-states (PDOS) for the ideal PbTe and L1$_1$ ordered AgSbTe$_2$ phases[38] are shown in Fig. 4. We find that the AgSbTe$_2$ compound is dynamically stable (i.e., all phonon frequencies are real) and that its average vibrational frequencies are higher than those of PbTe, consistent with the lighter average atomic mass of the former. The most pronounced differences are seen in the acoustic region, where both the average sound velocity (as measured by the curvature of the PDOS near ω=0) and the partial mode Te character are smaller in PbTe than in AgSbTe$_2$. Accordingly, in PbTe the Te ions constitute the optical frequency peak centered around 95 cm$^{-1}$, while the Pb ions dominate in the acustic region. In AgSbTe$_2$, due to the closer similarity of the ionic masses, the vibrational mode character is roughly evenly distributed between the cations and anions at both acoustic and optic frequencies. We conclude that several effects may combine to cause the significant lowering of thermal conductivity observed in LAST alloys: (i) atomic-scale roughness of the precipitate-matrix interface shown in Fig. 3, (ii) mismatch of the acoustic velocities

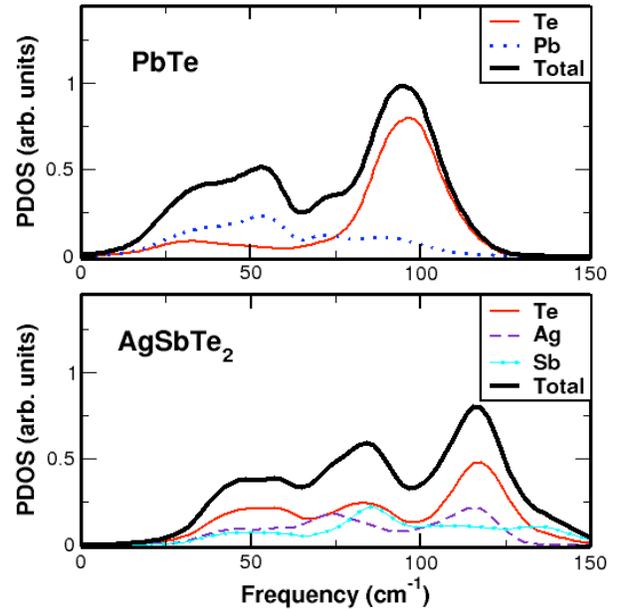

**Figure 4 (color online).** Phonon densities of states (PDOS) in rocksalt PbTe and L1$_1$ AgSbTe$_2$, and the partial contributions of each constituent element to the PDOS.

between the precipitates and the matrix and (iii) mismatch of the ionic character of the vibrational modes, causing bending and scattering of acoustic waves.

S.B. and V.O. gratefully acknowledge financial support from NSF under grant No. DMR-0427638 and from the FCRP Focus Center for Functional Engineered Nano Architectonics.